\documentclass{article}
\usepackage{spconf,amsmath,graphicx}
\usepackage{booktabs}
\usepackage{amssymb}
\usepackage{cite}
\usepackage{xurl}
\usepackage[urlcolor=blue]{hyperref}

\begin{document}
\ninept

%

\title{Disentangling content and fine-grained prosody information via hybrid ASR bottleneck features for voice conversion}




%

\name{Xintao Zhao$^{1, \dagger}$,
\thanks{$\dagger$ Work performed while interning at Huya Inc.}
        Feng Liu$^2$,
        Changhe Song$^1$,
       Zhiyong Wu$^{1,3,\ddagger}$
\thanks{
$\ddagger$ Corresponding author.},
 Shiyin Kang$^{2,\ddagger}$,
 Deyi Tuo$^2$,
 Helen Meng$^{1,3}$}

\address{Author Affiliation(s)}

\address{
   \\
    $^1$ Tsinghua-CUHK Joint Research Center for Media Sciences, Technologies and Systems, \\
Shenzhen International Graduate School, Tsinghua University, Shenzhen, China\\
$^2$ Huya Inc, Guangzhou, China\\
$^3$ Department of Systems Engineering and Engineering Management, \\ The Chinese University of Hong Kong, Hong Kong SAR, China\\
    \small{\{zxt20,	sch19\}@mails.tsinghua.edu.cn,
    \{liufeng1, kangshiyin, tuodeyi\}@huya.com,
	\{zywu,	hmmeng\}@se.cuhk.edu.hk}
    }

\maketitle


\maketitle

\begin{abstract}
Non-parallel data voice conversion (VC) have achieved considerable breakthroughs recently
through introducing
bottleneck features (BNFs) extracted by the automatic speech recognition(ASR) model.
However, 
selection of BNFs
have a significant impact on VC result. For example, when extracting BNFs from ASR trained with Cross Entropy loss (CE-BNFs) and feeding into neural network to train a VC system, the timbre similarity of converted speech is significantly degraded. 
If BNFs are extracted from ASR trained using Connectionist Temporal Classification loss (CTC-BNFs), the naturalness of the 
converted speech may decrease. 
This phenomenon is caused by the difference of information contained in BNFs.
%
%
In this paper, we proposed an any-to-one VC method using hybrid 
bottleneck features extracted from 
CTC-BNFs and CE-BNFs to complement each other advantages. 
 Gradient reversal layer
and instance normalization 
were used to extract prosody information from CE-BNFs and content information from CTC-BNFs. Auto-regressive decoder and Hifi-GAN vocoder were used to generate high-quality waveform.
Experimental results show that our proposed method achieves higher similarity, naturalness, quality than baseline method 
and reveals the differences between the information contained in CE-BNFs and CTC-BNFs as well as the influence they have on the converted speech.


\end{abstract}




%

\section{Introduction}
\label{sec:intro}

Voice conversion (VC) aims at converting a speech uttered by source speaker as if it was uttered by a target speaker, while maintaining speech content. 
Conventional VC systems require paired training utterances 
which are expensive in practice.
Non-parallel VC is a more challenging but more practical task in applications. The training utterances 
are only recorded by the target speaker, which is easier for collection than parallel data.
A common way to achieve non-parallel VC is disentangling the linguistic and non-linguistic information carried by the source and target utterances respectively, then combining them and synthesizing the converted utterance. 
In order to generate waveform with high quality and similarity, 
it is crucial to keep these information accurate and 
disentangled as much as possible.
Some methods tried to 
disentangle content information with timbre and pitch in an unsupervised manner
\cite{hsu2017voice,qian2020unsupervised,wang2021adversarially}. 
However, due to the difficulty of the unsupervised learning, the quality and intelligibility of converted waveform inevitably degrades.

\begin{figure}[t]
  \centering
  \includegraphics[width=1.0\linewidth]{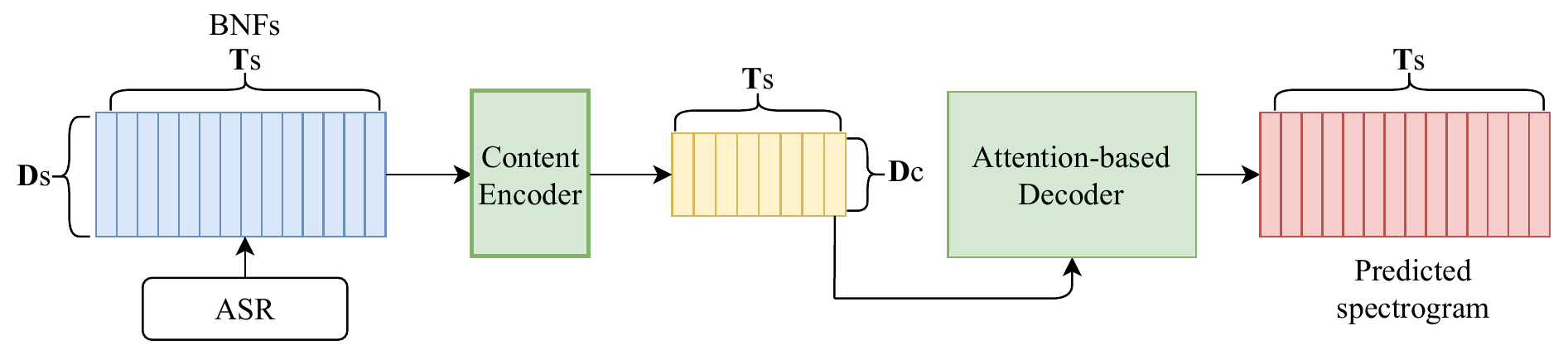}
  \caption{Baseline model architecture}
  \label{fig:baseline_img}
\end{figure}

Recognition-synthesis based VC is another representative approach of the non-parallel VC
and has been shown to achieve noticeable performance improvement\cite{sun2016phonetic,ding2020improving,zhang2021non,liu2021any}. 
Such methods provide more robust linguistic representations based on the knowledge from an automatic speech recognition (ASR) model trained with huge amount of data.
Therefore, information contained in the representations is considered to be highly related to the content information 
and disentangled with speaker-related information to a certain degree\cite{liu2021any}.
An intermediate embedding is extracted from a certain layer of ASR model, either phonetic posteriorgrams (PPGs)\cite{sun2016phonetic} from the last layer regardless of preformed extractor structure, or bottleneck features (BNFs) \cite{wang2021accent} from the penultimate layer with well-designed dimension, and then fed into a VC system. 
BNFs based VC framework, as shown in Fig. \ref{fig:baseline_img} is proven to be more robust when facing recognition error than PPGs in practice. 

%
%
\begin{figure*}[t]
  \centering
  \includegraphics[width=1.0\linewidth]{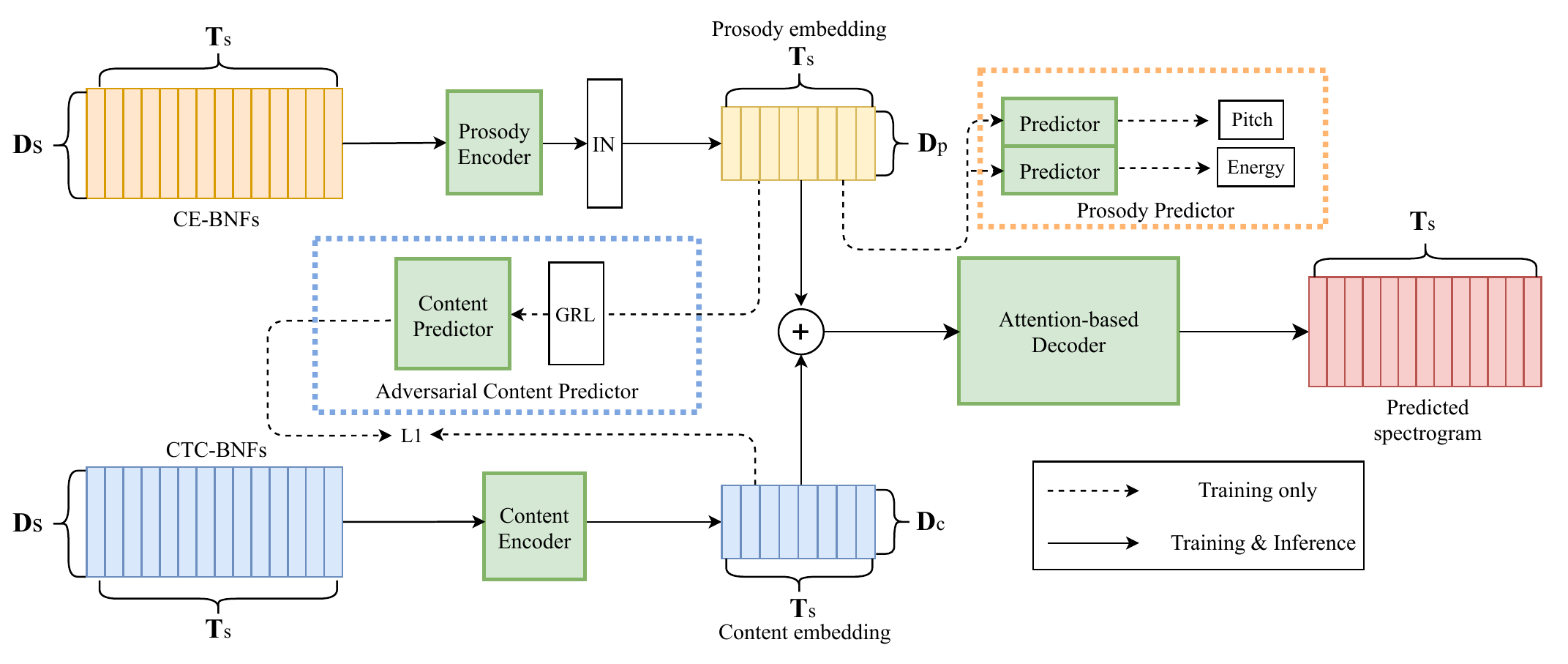}
  \caption{Proposed model architecture.}
  \label{fig:proposed_img}
\end{figure*}
Prior approaches in recognition-synthesis based VC 
extract BNFs 
from ASR model trained with cross entropy loss (CE loss) 
on the time-aligned training data (CE-ASR) 
and illustrate noticeable performance\cite{zhang2019non,zhang2019sequence}. However, we find that
the BNFs extracted by CE-ASR (CE-BNFs) contain extra prosody information and timbre information that may leak into the voice conversion task\cite{ding2020improving}.
Though 
the converted waveform could get fine-grained prosody variation that is
similar to the source utterance
and has high naturalness, the similarity to target speaker and speech quality
degrade significantly due to the preserved redundant source timbre information in CE-BNFs. 
A PPGs-based approach based on the ASR model trained with connectionist temporal classification loss (CTC-ASR)\cite{graves2006connectionist} is proposed recently \cite{zhang2021non}. By only preserving the non-blank-token frames of the PPGs based on the peaky behavior of the CTC models, a pure linguistic representation is extracted. 
Considering that for utterances with same content but spoken by different speakers, the diversity of converted speeches may degrade if only non-blank-token frames are fed into VC system.
Inspired by the influence of PPGs with different training losses on VC mentioned in \cite{zhang2021non}, we 
studied the impact of BNFs with different ASR training losses to VC and find that 
the BNFs 
extracted from CTC-ASR (CTC-BNFs) 
preserve less non-linguistic information 
indicating better disentangling characteristics. 
Therefore, 
the converted waveform has limited naturalness but higher target speaker similarity and quality due to the
elimination
of prosody and timbre information from source speaker.
It is easy to find that CE-BNFs and CTC-BNFs can 
complement each other.
In this paper, 
to retain the good disentangling characteristics of CTC-BNFs, and meanwhile preserve the fine-grained prosody information contained in CE-BNFs, 
we proposed an any-to-one VC system using hybrid 
BNFs extracted by two ASR systems
with CTC and CE loss respectively. 
The proposed approach attempted to extract prosody information from CE-BNFs, and content information from CTC-BNFs. 
Compared to mel-spectrogram, CE-BNFs has more stable data distribution on diversified input waveform and is more suitable for extracting prosody embedding in our any-to-one VC task. 
On the other hand, the proposed model tried to extract content embedding from CTC-BNFs rather than CTC-PPGs because BNFs are more robust
to recognition error.
The contributions of our work include:
\begin{itemize}
    \item[1)] We proposed a VC method by virtue of hybrid BNFs from CTC-BNFs and CE-BNFs. This method
    leverages the complementary information of the two BNFs
    and achieved high similarity, naturalness and quality of the converted speech.
    \item[2)] We demonstrated the information differences contained in CTC-BNFs and CE-BNFs through appropriate experiments.
    \item[3)] An adversarial-prediction structure was proposed to disentangle content and prosody information.
    \item[4)] Experimental results revealed that BNFs are more robust to ASR recognition errors than PPGs for VC.
\end{itemize} 
\section{Methodology}
\label{sec:method}


The proposed any-to-one VC system, shown in Fig.\ref{fig:proposed_img}, consists of five components: 
(i) A prosody encoder followed by instance normalization (IN)\cite{chou2019one} to encode CE-BNFs into prosody embedding. 
(ii) A content encoder to encode CTC-BNFs into content embedding. 
(iii) An adversarial content predictor with gradient reversal layer (GRL)\cite{ganin2016domain} 
to reduce the information overlapping between prosody embedding and content embedding. 
(iv) A prosody predictor consists of two sub-predictors 
to predict pitch and energy 
so as to force the prosody encoder work correctly.
(v) An auto-regressive decoder with location-sensitive attention\cite{chorowski2015attention} for high-quality spectrogram
generation. 
Although (iii) and (iv) increase the complexity 
at training stage, the increase in the number of parameters is 
acceptable as we do not need to load their parameters at inference stage.

\subsection{ASR system}
The very recent conformer\cite{gulati2020conformer} based ASR acoustic model achieved state-of-the-art recognition performance due to the combination of Transformer and CNN models. The conformer models are good at capturing content-based global interactions as well as local features. We trained two ASR models with conformer encoder but different loss functions: CE loss and CTC loss. We extract 256-dims BNFs from the last hidden layer of conformer encoder.

\subsection{Encoder-Decoder model}
As shown in Fig.\ref{fig:proposed_img}, the proposed conversion model is an encoder-decoder architecture, following the setting of Tacotron2\cite{shen2018natural}. The input CTC-BNFs were passed through a content encoder firstly. The content encoder consists of a stack of 2 fully-connected layers and a set of 1D convolutional layers, followed by a multi-layer highway network\cite{srivastava2015highway} and a bidirectional GRU layer.
The outputs of content encoder were then concatenated with prosody embedding extracted by prosody encoder.
The concatenated content and prosody embeddings were further 
fed into the auto-regressive decoder
to generate high-quality spectrogram. 

\subsection{Prosody information extraction}
Prosody encoder consists of a Conv1d layer and a GroupNorm layer followed by a BLSTM structure. 
The prosody embedding has a significant narrow dimension compared to CE-BNFs to restrict the information passed by.
To further eliminate the timbre information, an IN module was introduced here.
The output of prosody encoder was fed into adversarial content predictor, prosody predictor and decoder at the same time. 
Each sub-predictor in prosody predictor consists of 3 fully-connected layers followed by GeLU activation \cite{hendrycks2016gaussian} and layer normalization\cite{ba2016layer}.
The prosody loss is defined as:
\begin{equation}
\small
F_{prosody} = IN(E_{prosody}(BNF_{ce}))
\label{eq:prosody_embedding}
\end{equation}
\begin{equation}
\small
\hat{E},\hat{P} = P_{prosody}(F_{prosody})
\label{eq:predictor_PE}
\end{equation}
\begin{equation}
\small
{L}_{energy} = \|E-\hat{E}\|_{1}
\label{eq:energy}
\end{equation}
\begin{equation}
\small
{L}_{pitch} = \|P-\hat{P}\|_{1}
\label{eq:pitch}
\end{equation}
\begin{equation}
\small
{L}_{prosody} = \lambda _{p} * L_{pitch} + \lambda _{e} *L_{energy} 
\label{eq:pitch}
\end{equation}
where $BNF_{ce}$ is the input CE-BNFs, $F_{prosody}$ is the prosody embedding, $E_{prosody}$ and $P_{prosody}$ correspond to prosody encoder and predictor. $P$, $\hat P$ are the ground truth and predicted pitch, $E$, $\hat E$ are the ground truth and predicted energy respectively. ${L}_{prosody}$ is the total prosody loss, where we simply set $\lambda_{p}$ and $\lambda_{e}$ to 1.

\subsection{Adversarial learning for information disentanglement}\label{subsection_Adv_predictor}

An adversarial content predictor network inspired by \cite{wang2021adversarially} was designed to reduce the overlap of information between prosody embedding and content embedding. The network consists of a GRL and a prediction head layer\cite{liu2020mockingjay} as content predictor.
The output of the adversarial content predictor network has the same size as the content embedding extracted by the content encoder. The gradient of the adversarial network is reversed by 
GRL before backward propagated to the prosody encoder. 
The adversarial loss is defined as:

\begin{equation}
F_{content} = E_{content}(BNF_{ctc})
\end{equation}
\begin{equation}
\hat{F}_{content} = P_{content}(GRL(F_{prosody}))
\end{equation}
\begin{equation}
\small
\label{eq:adv}
{L}_{adv} = \left|\left|\hat{F}_{content} - F_{content}\right|\right|_{1}
\end{equation}
where 
$BNF_{ctc}$ is the input CTC-BNFs of content encoder $E_{content}$,
$F_{content}$ and $\hat{F}_{content}$ correspond to the content embedding generated by content encoder and the output of adversarial content predictor. $P_{content}$ and $GRL$ means 
content predictor
and gradient reversal layer, respectively. The optimization of $L_{adv}$ forces 
prosody embedding
to contain as little 
overlapped content
information as possible because of the reversed gradient.


\subsection{Voice conversion loss}

We computed the reconstruction loss between the input and reconstructed mel spectrograms.
To speed up the training process, we generated a diagonal matrix and calculate the alignment loss between the matrix and alignment of decoder as $L_{align}$.
The reconstruction loss and the final objective loss functions are given as follows:
\begin{equation}
\small
\label{eq:recon}
{L}_{rec} = \|S-\hat{S}\|_{1}
\end{equation}
\begin{equation}
\small
\label{eq:loss}
{Loss} = \lambda _{adv} * L_{adv} + \lambda _{rec} * L_{rec} \\
+ \lambda _{a} *L_{align} + {L}_{prosody}
\end{equation}
where $S$ and $\hat S$ are the mel spectrograms of the ground truth and reconstructed speeches respectively, and $\lambda$s
are the loss weights for all the losses. 

Finally, Hifi-GAN \cite{kong2020hifi} was used as the vocoder to get high fidelity waveform and fast decoding speed.

\begin{figure}[t]
  \setlength{\belowcaptionskip}{-0.em}
  
  \centering
  \includegraphics[width=1.0\linewidth]{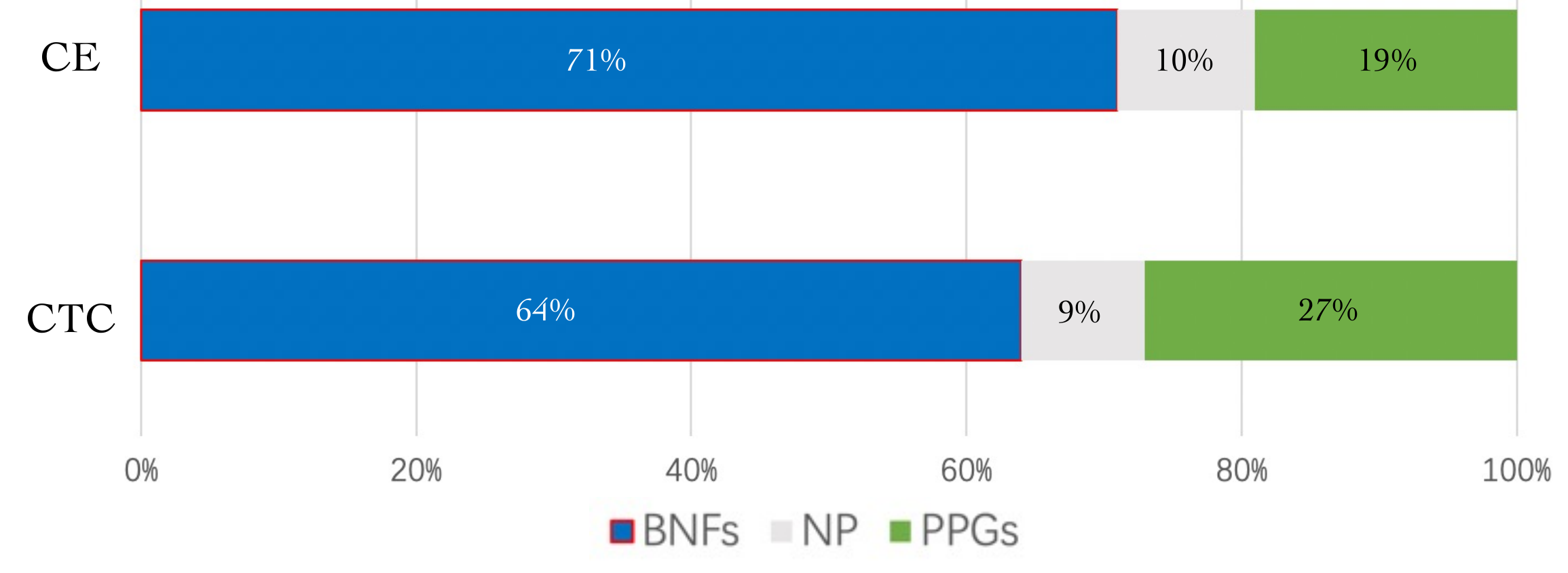}
  \caption{Results of the ABX-test for \textbf{Naturalness}. BNFs and PPGs correspond to the input of the baseline VC model. CE and CTC indicate the loss which is used in training ASR system. NP means no preference}
  \label{fig:ABX_BN_PPG_img}
  
\end{figure}


%
\section{Experiment}
\label{sec:typestyle}
\subsection{Training setup}
The voice conversion experiments were performed on a homegrown Mandarin corpus uttered by a female speaker. The original corpus duration is 2 hours in total. The training data is augmented by changing the speaking rate from 0.9 to 1.5 times to enhance the prosody diversity. ASR systems were trained with nearly 3000 hours Mandarin corpus uttered in live streaming scene, which may contain noisy and accented speech. We extract 256-dim BNFs and 331-dim PPGs from ASR model separately in our experiments.

We evaluated the speaker information contained in BNFs extracted from different ASR systems via a speaker classification network. 
We introduced a ResNet-32\cite{he2016deep} based neural network here. It would randomly select 2s bottleneck frames from each utterance as input and predict which speaker these frames belong to. 
Nearly 300 hours of Mandarin corpus from AISHELL-1\cite{aishell_2017}, AISHELL-3\cite{AISHELL-3_2020}, THCHS\cite{DBLP:journals/corr/WangZ15e} and VC corpus were used in training. The corpus contains 381 female speakers and 252 male speakers. 20 sentences were randomly chosen from each speaker as the validation set.

All the audios were down-sampled to 16KHz. Mel spectrograms were computed through a short time Fourier transform (STFT) using a 50ms frame size, 10ms frame hop and a Hann window function. Fbank features were used in ASR system with 25ms frame size and 10ms frame hop. Our ASR systems were based on the implement of WeNet, 
 an E2E speech recognition toolkit\footnote{\href{https://github.com/wenet-e2e/wenet}{https://github.com/wenet-e2e/wenet}}.
A pretrained HiFi-GAN \cite{kong2020hifi} vocoder was used to synthesize the audios from the mel spectrogram.
Baseline and proposed systems were evaluated under the same settings, otherwise mentioned.
Demo page is available\footnote{\href{https://thuhcsi.github.io/icassp2022-hybrid-bottleneck-vc/}{https://thuhcsi.github.io/icassp2022-hybrid-bottleneck-vc/}}.


\subsection{Subjective evaluation}


\begin{figure}[t]
  \setlength{\belowcaptionskip}{-10cm} 
  \centering
  \includegraphics[width=1.0\linewidth]{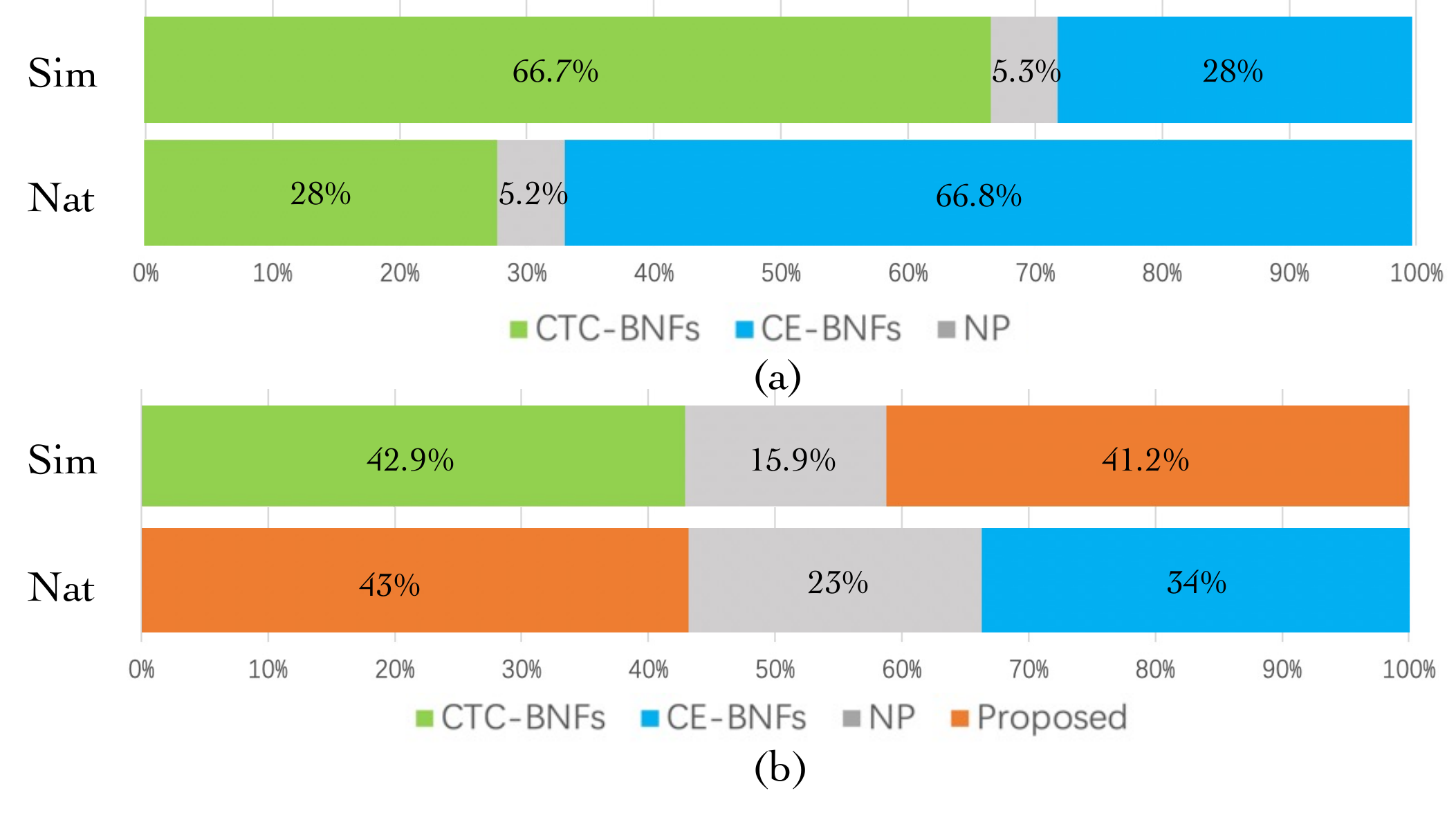}
  \caption{Result of the ABX test for \textbf{Naturalness} and \textbf{Similarity}.
  Where \textbf{*-BNFs} means baseline model with different BNFs as input.
  }
  \label{fig:ABX_CE_CEC_Proposed_img}
\end{figure}

In order to show the impact of different linguistic features on the VC system, we conducted an ABX test to compare the 
naturalness or the timbre similarity 
of the converted speech. 
“A” and “B” were converted speech
samples obtained with different methods
and “X” was a real speech sample obtained from a source speaker in naturalness evaluation which has correct and natural prosody, or a real speech sample from the target speaker in similarity evaluation. 
Each listener was presented 20 utterances for each ABX pair. 
The result in Fig.\ref{fig:ABX_BN_PPG_img} shows that BNFs are more robust than PPGs in terms of recognition errors, which lead to significant improvement of naturalness no matter which loss function are used in training the ASR system. 
As the results in Fig.\ref{fig:ABX_CE_CEC_Proposed_img} (a) reveals, CE-BNFs get lower similarity but higher naturalness, whereas CTC-BNFs get the adverse result.  Fig.\ref{fig:ABX_CE_CEC_Proposed_img} (b) shows the proposed method 
combined the merits
of the two types of BNFs and generated speech with high similarity and naturalness.

\begin{table}[h]
    \vspace{-1.0 em}
    \setlength\abovedisplayskip{0pt}
	\caption{
    Speaker classification accuracy and Cosine similarity calculated by CTC-BNFs and CE-BNFs. Cosine similarity stands for the average cosine similarity of all neighbouring BNF vectors.
	}
	\label{tab:cosine}
	\centering

	\begin{tabular}{lccc}
		\toprule
		 &\small{CTC-BNFs} &\small{CE-BNFs} \\
        \midrule
        
		\small{Cosine similarity} &0.791 &\textbf{0.769}  \\
		\small{Speaker classification accuracy} &\textbf{32.2\%} &68.89\%   \\
		\bottomrule
	\end{tabular}
\end{table}

\subsection{Objective evaluation}

%
%
Table \ref{tab:cosine} shows the information difference between CTC-BNFs and CE-BNFs. \textit{Speaker cls acc} indicates the accuracy of our speaker classification task using different BNFs as input. It is shown that CE-BNFs have significantly high accuracy compared to CTC-BNFs, this indicates CE-BNFs preserve much more timbre information. 
We then calculated average \textit{cosine similarity} between neighboring BNF vectors.  
Provided that the neighbouring BNF vectors mostly belong to the same phoneme,
lower cosine similarity 
means more prosodic variation and richer prosody information contained in BNFs. 

Table \ref{tab:Objct} shows the Mel-cepstral distortion (MCD), root-mean-square error (RMSE) in F0 and energy. MCD was used to measure how close the converted is to the target speech. 
M2VoC parallel data\footnote{\href{http://challenge.ai.iqiyi.com/detail?raceId=5fb2688224954e0b48431fe0}{http://challenge.ai.iqiyi.com/detail?raceId=5fb2688224954e0b48431fe0}} were used for MCD calculation. We adopted Pysptk tools to extract the MCCs and dynamic time warping (DTW) to align the target and converted speech parameters. Lower value indicates higher similarity. It can be observed that CE-BNFs-VC system gets higher MCD 
than CTC-BNFs-VC system because the redundant timbre information contained in CE-BNFs limited similarity and quality.
Energy and F0 RMSEs were used to measure the naturalness of the converted waveform. Both F0 and energy sequences were performed min-max normalization before calculation. We calculated RMSE between converted and source waveform because source waveform was uttered by real speakers with fine-grained prosody and high naturalness. It is found that CE-BNFs-VC system achieves lower energy and F0 RMSEs because the extra fine-grained prosody information preserved contributes to high naturalness. 
Overall, the proposed method outperforms two baseline methods in all metrics.

\begin{table}[h]
    \vspace{-1.0 em}
	\caption{
	MCD, Energy RMSE and F0 RMSE calculated between the waveforms generated by different systems.
 	Energy 
 	and F0 RMSEs were calculated between \textbf{source} and \textbf{converted} waveform while MCD was calculated between 
 	\textbf{target} and \textbf{converted} waveform.
	}
	\label{tab:Objct}
	\centering

	\begin{tabular}{lccc}
		\toprule
		 &\small{MCD} &\small{Energy RMSE} &\small{F0 RMSE}\\
        \midrule
        
		\small{CTC-BNFs-VC system} &7.494 &0.306  &0.395 \\ 
        \small{CE-BNFs-VC system} &7.760 &0.298 &0.365 \\ 
        \small{Proposed-VC system} &\textbf{7.453} &\textbf{0.286} &\textbf{0.358} \\ 
		\bottomrule
	\end{tabular}
\end{table}

\subsection{Ablation study}
To understand how much each component contributes to the model, %
mean opinion scores (MOS) tests were conducted to evaluate the speech quality, prosody naturalness and timbre similarity of the converted waveform. Each listener was presented 25 utterances for each experiment,
and was asked to give a 5-point scale score with 1 point interval.
Table \ref{tab:ablation} shows that all components have significant impact on our proposed model. Model without 
IN leads to significantly performance degradation in similarity and quality because IN could eliminate timbre information of source speaker.
The model without prosody predictor leads to degradation in naturalness. Adversarial learning also plays an important role in disentanglement. If we replace CE-BNFs by mel-spectrogram like \cite{karlapati2020copycat}, due to the variation of source waveform, the prosody extraction will
fail
in our experiment.

\begin{table}[h]
    \vspace{-1.0 em}
	\caption{
	The MOS with 95\% confidence intervals shows the impact of different modules in network on naturalness, similarity and quality. \textbf{Mel-Prosody encoder} row is the result gained by feeding mel-spectrogram into prosody encoder to extract prosody embedding.
	}
	\label{tab:ablation}
	\centering

	\begin{tabular}{lccc}
		\toprule
		 &\small{Naturalness} 
		 &\small{Similarity} 
		 &\small{Quality}\\
        \midrule
        
		\small{Proposed} 
		&3.65(-0.00)
		&3.75(-0.00)  
		&3.96(-0.00) \\ 
		
        \small{w/o IN} 
        &3.16(-0.49)
        &2.89\textbf{(-0.86)}
        &2.90\textbf{(-1.06)}\\ 
        
        \small{w/o Prosody predictor} 
        &3.02\textbf{(-0.65)}
        &3.19(-0.56)
        &3.23(-0.73) \\
        
        \small{w/o Adv predictor} 
        &3.20(-0.45) 
        &3.36(-0.39) 
        &3.58(-0.38) \\ 
        \bottomrule
        
        \small{Mel-Prosody encoder} 
        &2.17(-1.45)
        &2.31(-1.44)
        &2.76(-1.20)\\ 
        
		\bottomrule
	\end{tabular}
\end{table}
\section{Conclusion}
\vspace{-1em}
\label{sec:majhead}
\setlength{\parskip}{0.pt}

In this paper, we proposed an any-to-one voice conversion model using hybrid BNFs. 
The proposed method retains the good disentangling characteristics of CTC-BNFs while preserving the fine-grained prosody information contained in CE-BNFs, leading to superior performance than the baseline models in naturalness and similarity. 
The experiments demonstrate that CE-BNFs contains much more timbre and prosody information than CTC-BNFs. CE-BNFs is a better choice when attempts are made to extract prosody information from various source waveform than the mel-spectrogram. Other than that, our ABX test also proves that BNFs features are more robust choice than PPGs in terms of recognition error. In the future, we will focus on any-to-many VC task and try to extract prosody information and content information separately from two different waveforms to further control the prosody of any content.


\textbf{Acknowledgement}: This work is supported by National Natural Science Foundation of China (NSFC) (62076144) and National Social Science Foundation of China (NSSF) (13\&ZD189)



\bibliographystyle{IEEEbib}
\bibliography{content/references}

\end{document}